\documentclass[pre,twocolumn,amsmath,amssymb,showpacs]{revtex4}
\usepackage{graphicx}
\usepackage{dcolumn}
\usepackage{bm}
\usepackage[usenames]{color}
\usepackage[all]{xy}
\usepackage{amsmath}
\usepackage{natbib}
\usepackage[toc]{appendix}
\usepackage[usenames]{color}

\def \bea{\begin{eqnarray}}
\def \eea{\end{eqnarray}}

\begin{document}
\bibliographystyle{plainnat}
\newcommand{\dd}{\text{d}}
\newcommand{\ee}{\text{e}}
\newcommand{\kb}{k_\text{\tiny B}}
\newcommand{\td}{\tau_\text{r}}
\newcommand{\fa}{f_\text{\tiny A}}
\newcommand{\Ta}{T_\text{\tiny A}}
\newcommand{\ti}{t_\text{i}}
\newcommand{\tf}{t_\text{f}}

\providecommand{\avg}[1]{\left \langle #1 \right \rangle}
\providecommand{\pnt}[1]{\left ( #1 \right)}
\providecommand{\brt}[1]{\left [ #1 \right]}

\title{Modeling the dynamics of a tracer particle in an elastic active gel}
\author{E. Ben Isaac$^1$, \'E. Fodor$^2$, P. Visco$^2$, F. van Wijland$^2$ and Nir S. Gov$^1$}
\affiliation{ $^1$ Department of Chemical Physics, The Weizmann Institute
of Science, Rehovot 76100, Israel \\
$^2$ Laboratoire Mati\`ere et Syst\`emes Complexes, UMR  7057 CNRS/P7, Universit\'e Paris Diderot, 10 rue Alice Domon et L\'eonie Duquet, 75205 Paris cedex 13, France}

\begin{abstract}
The internal dynamics of active gels, both in artificial (in-vitro) model systems and inside the cytoskeleton of living cells, has been extensively studied by experiments of recent years. These dynamics are probed using tracer particles embedded in the network of biopolymers together with molecular motors, and distinct non-thermal behavior is observed. We present a theoretical model of the dynamics of a trapped active particle, which allows us to quantify the deviations from equilibrium behavior, using both analytic and numerical calculations. We map the different regimes of dynamics in this system, and highlight the different manifestations of activity: breakdown of the virial theorem and equipartition, different elasticity-dependent "effective temperatures" and distinct non-Gaussian distributions. Our results shed light on puzzling observations in active gel experiments, and provide physical interpretation of existing observations, as well as predictions for future studies.

\end{abstract}


\maketitle

\section{Introduction}

In-vitro experiments have probed the
non-thermal (active) fluctuations in an "active gel", which is most commonly realized as a
network composed of cross-linked filaments (such as actin) and molecular motors (such as myosin-II) \cite{Schmidt,marina,Mackintosh,MackintoshNT}. The fluctuations
inside the active gel were measured using the tracking of individual
tracer particles, and used to demonstrate the active
(non-equilibrium) nature of these systems through the breaking of
the Fluctuation-Dissipation theorem (FDT) \cite{Mackintosh}. In
these active gels, myosin-II molecular motors generate relative
motion between the actin filaments, through consumption of ATP, and
thus drive the athermal random motion of the probe particles
dispersed throughout the network. This tracking technique was also
implemented in living cells \cite{elbaum,fredberg,ahmed}. The motion of
these tracers in cells was also shown to deviate from simple thermal
Brownian diffusion.

There are several puzzling observations of the dynamics of the tracer particles inside the active gels, for example, the distinct non-Gaussianity of the displacement correlations and their time-dependence \cite{Schmidt,Koenderink2012pre,konderink2014}. We propose here a simple model for the random active motion of a tracer particle within a (linearly) elastic active gel, and use our model to resolve their distinct non-equilibrium dynamics. On long time-scales the tracer particles are observed to perform hopping-like diffusion, which is beyond the regime of the present model, and will be treated in a following work, as will be the introduction of non-linear elasticity \cite{nonlinear}.
The activity is modeled through colored shot-noise \cite{Annunziato,eyalPRL} and the elastic gel is described by a confining harmonic potential.
We use the model to derive expressions directly related to the experimentally-accessible observations: such as the position and velocity distributions and their deviations from the thermal Gaussian form. Our model allows us to offer a physical interpretation to existing experiments, to characterize the microscopic active processes in the active-gel, and to make specific predictions for future exploration of the limits of the active forces and elasticity.
The simplicity of this model makes this model applicable to a wide range of systems, and allows us to gain analytic solutions, intuition and understanding of the dynamics, which is usually lacking in out-of-equilibrium systems. This would be more difficult to obtain with more complex description of the gel, such as visco-elastic that has more intrinsic time-scales.

\section{Model}

Our model treats a particle in a harmonic potential,
kicked randomly by thermal and active forces (active noise) \cite{eyalPRL}. The
corresponding Langevin equation for the particle velocity $v$ (in
one dimension or one component in higher dimensions, with the mass
set to $m=1$)
\begin{eqnarray}
\dot{v}&=&-\lambda v +f_a+f_T-\frac{\partial U(x)}{\partial x}
\label{vdot}
\end{eqnarray}
where $\lambda$ is the effective friction coefficient and the
harmonic potential is: $U(x)=kx^2/2$, with $k$ proportional to the
bulk modulus of the gel (related to the gel density, cross-linker
density and other structural factors). The thermal force $f_T$ is an
uncorrelated Gaussian white noise: $\langle f_T(t)f_T(t')
\rangle=2\lambda T\delta(t-t')$, with $T$ the ambient
temperature, and Boltzmann's constant set to $k_B=1$.

We model the active force $f_a$ as arising from the independent action of
$N_m$ molecular motors, each motor producing pulses of a given fixed force
$\pm f_0$, for a duration $\Delta\tau$ (either a constant or drawn from
a Poissonian process with an average value $\Delta \tau$, i.e. shot-noise), with a random direction (sign). The
active pulses turn on randomly as a Poisson process with an average
waiting time $\tau$ (during which the active force is zero), which determines the "duty-ratio" of the motor
(the probability to be turned "on"):
$p_{on}=\Delta\tau/(\tau+\Delta\tau)$.

\section{Results: mean kinetic and potential energies}

The mean-square velocity and position fluctuations of the trapped particle, essentially the mean kinetic ($T_v=\langle v^2\rangle$) and potential ($T_x=k\langle x^2\rangle$) energies, can be calculated for the case of shot-noise force correlations (details given in the Appendix, Eqs.\ref{A1}-\ref{tvapprox} and Figs.\ref{figx1}-\ref{simSquare}). Note that the mean $\langle\cdot\rangle$ is over many realizations of the system, or over a long time. In the limit of vanishing trapping potential the position fluctuations $\langle x^2\rangle$ diverge, but the potential energy approaches a constant: $T_x|_{k\rightarrow0}\rightarrow f_0^2\Delta\tau/\lambda$. The kinetic energy approaches the constant value for a free particle \cite{eyalPRL}: $T_v|_{k\rightarrow0}\rightarrow T_x/(1+\lambda\Delta\tau)$. We therefore find that the virial theorem is in general not satisfied in this active system, which in a harmonic potential gives $T_v|_{eq}=T_x|_{eq}$, even in the limit of weak trapping. The virial theorem, and equipartition, breaks down due to the strong correlations between the particle position and the applied active force: In the limit of perfect correlations, the particle is stationary at $x=\pm x_0$ when the force is turned on (the stationary position in the trap where the potential balances the active force: $x_0=f_0/k$), and at $x=0$ when it is off. In this extreme case the potential energy is finite while the kinetic energy is zero.

In the limit of strong trapping $k\rightarrow\infty, k/\lambda^2\gg1$, the potential energy behaves as: $T_x\propto k^{-1}$ (Eqs.\ref{txweak},\ref{txA}), while the kinetic energy decays faster as: $T_v\propto k^{-3/2}$ (Eq.\ref{tvapproxfar1}). One can understand this limit as follows: When the trapping is very strong, the shortest time-scale in the problem is the natural oscillation frequency in the trap, $\omega_k\sim\sqrt{k}$. In this regime of $k\Delta\tau^2\gg1$ we find that during the active pulse $\Delta\tau$, the particle reaches $x_0$, and the mean potential energy is therefore proportional to $T_x\sim kx_0^2\propto1/k$. The kinetic energy in this limit decays faster, since the fraction of time that the particle is moving is only during the acceleration phase determined by the time-scale $\omega_k^{-1}\sim\sqrt{k}$.
We therefore find that in the presence of strong elastic restoring forces the potential energy will be much larger than the kinetic energy, in an active system ($T_x\gg T_v$). This was recently found in the study of active semi-flexible polymers \cite{abhijit2014}.

Note that in a real active gel the different parameters maybe coupled: larger local density of the network filaments increases the local value of the elastic stiffness parameter $k$, but may also increase locally the density of motors and their ability to exert an effective force, thereby increasing $N_m$ and $f_0$. The tracer bead behavior as expressed by $T_v$ and $T_x$ can therefore be a complex function of the local network parameters.

\section{Results: Velocity and position distributions}

\begin{figure}
\includegraphics[width=0.6\columnwidth]{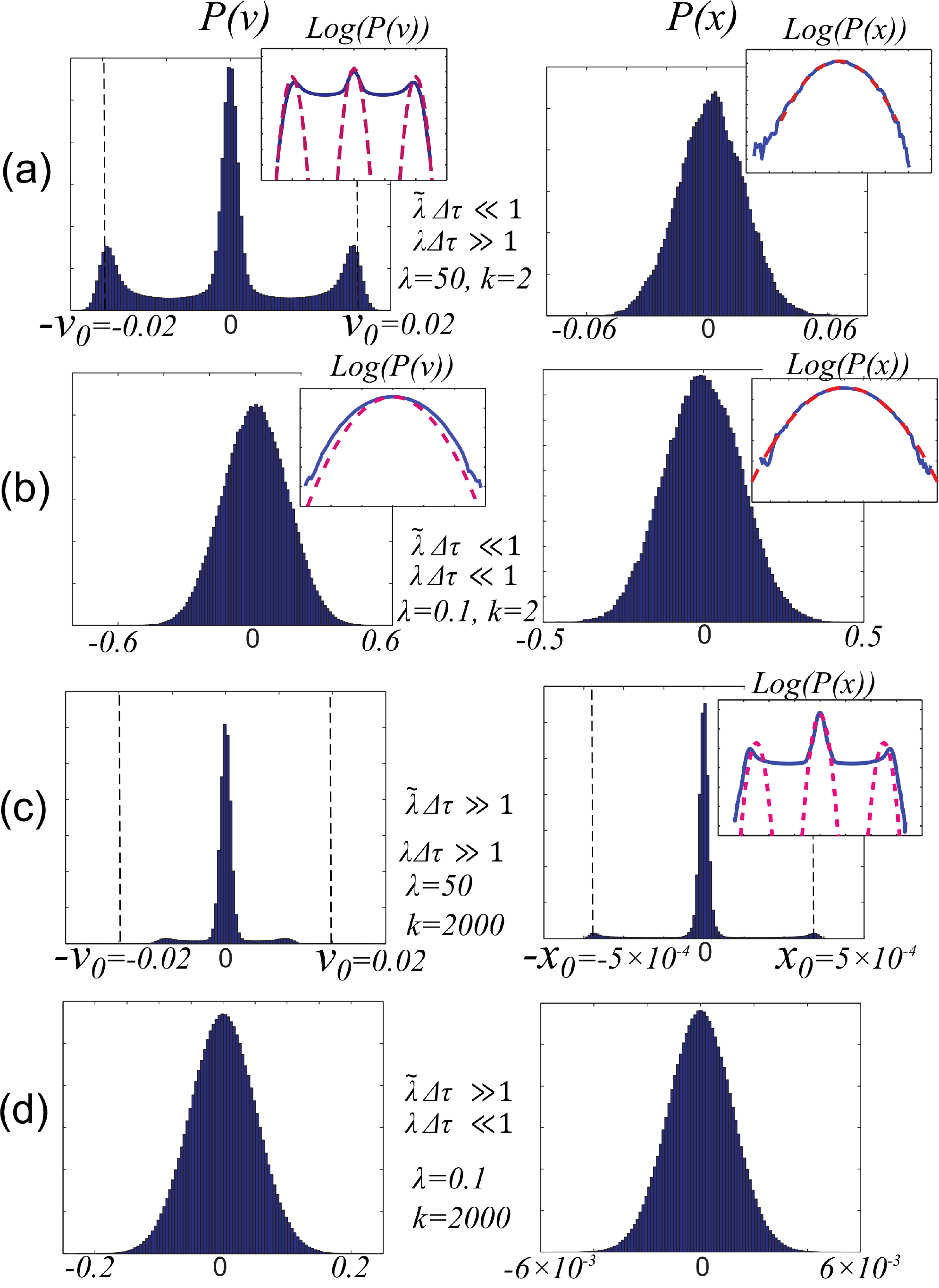}
\caption{Distribution of position and velocity for particle trapped within a harmonic trap of various stiffness ($k=1,1000$ for a,b and c,d respectively), and for different regimes of friction ($\lambda=50,0.1$ for a,c and b,d respectively). The time-scale of the active bursts is $\Delta\tau=0.1$, the amplitude of the active force $f_0=1$ and the waiting time $\tau=1$ (so that $p_{on}\approx0.1$). For simplicity we plot the behavior for the case of a single motor with a constant burst duration. The insets compare the simulated distribution (blue line) to the analytic approximation (red dashed line), in log-linear scale, as simple Gaussians or as a sum of shifted thermal Gaussians.}\label{Fig2}
\end{figure}

The distributions of the velocity and position in the different regimes are shown in Fig.\ref{Fig2}, for the case of a single active motor. The simulations of the model were carried out using explicit Euler integration of Eq.\ref{vdot} (see also the Appendix for details). We study this case in order to highlight the deviations from Gaussian (equilibrium-like) behavior, which is restored by many simultaneous motors \cite{eyalPRL}.
In an infinite gel, with a constant density of motors, we may therefore treat the distant (and numerous) motors as giving rise to an additional thermal-like contribution to the tracer dynamics (Eqs.\ref{tvapproxfar},\ref{tvapproxfar1}), while the non-equilibrium behavior is dominated by a single proximal motor \cite{Schmidt}.

In the limit of weak damping, $\lambda \Delta \tau\ll1$, both the position and velocity distributions are very close to Gaussian, with the width of the Gaussian distributions given by $T_v$ and $T_x$ (Fig.\ref{Fig2}b,d, Eqs.\ref{fullxsol},\ref{fullvsol}): $P(v)\propto\exp{(-v^2/(2 T_v))}$, $P(x)\propto\exp{(-kx^2/(2T_x))}$.

In the highly damped limit, $\lambda \Delta \tau\gg1$, the distributions become highly non-Gaussian (Fig.\ref{Fig2}a,c). We can make a useful approximation in this limit, by neglecting the inertial
term in Eq.(\ref{vdot}) and get the following equation for the
particle position $x$ inside the potential well
\begin{eqnarray}
\lambda v&=&-kx +f_a+f_T \label{vdotdamp} \\
\Rightarrow \dot{x}&=&-\tilde{\lambda}x+\frac{f_a+f_T}{\lambda}
\label{xeq}
\end{eqnarray}
where $\tilde{\lambda}=k/\lambda$. This equation is now analogous
to the equation for the velocity $v$ of a free particle
(Eq.(\ref{vdot}) when $U(x)=0$). Due to this analogy we can use the
analytic solutions for the free particle \cite{eyalPRL} to describe
the particle position in the well.
For weak trapping (Fig.\ref{Fig2}a), we therefore expect the position distribution to be roughly Gaussian, since we are in the limit of $\tilde{\lambda}\Delta\tau\ll1$ of Eq.(\ref{xeq}), with a width given by (from Eq.\ref{xeq},S6)
\begin{eqnarray}
T'_x&=&\frac{p_{on}N_m\lambda\left(\tilde{\lambda}\Delta\tau+e^{-\tilde{\lambda}\Delta\tau}-1\right)}{k^2\Delta\tau}f_0^2 \label{txt} \\
T_x&=&2\frac{p_{on}N_m\langle\Delta\tau\rangle}{\lambda\left(1+\tilde{\lambda}\langle\Delta\tau\rangle\right)}f_0^2\label{tx}
\end{eqnarray}
where $T'_x$ describes the case of a constant $\Delta\tau$ and $T_x$
the case of a Poissonian burst distribution, and fits well the calculated distribution (inset of Fig.\ref{Fig2}a). In the limit of weak confinement we expect the velocity distribution to approach the behavior of the free damped particle \cite{eyalPRL}, which is well approximated as a sum of thermal Gaussians, centered at $v=0,\pm v_0$ ($v_0=F_0/\lambda$). This is indeed a good approximation, as shown in the inset of Fig.\ref{Fig2}a.

For strong potentials ($\tilde{\lambda}\Delta\tau\gg1$, Fig.\ref{Fig2}c) we expect from
the analogy given in Eq.(\ref{xeq}) that the spatial distribution is now
well described by the sum of shifted thermal Gaussians (Fig.\ref{Fig2}c)\cite{eyalPRL},
centered at $x=0,\pm x_0$. The velocity distribution in this regime
is also non-Gaussian: the maximal active velocity is of order $v_0$ at the origin of the potential, but since the particle immediately slows due to the confinement (up to a complete stop at $\pm x_0$), the peaks of the distribution are located at roughly $\pm v_0/2$.

\section{Results: Non-Gaussianity of the displacement distribution}

\begin{figure}
\includegraphics[width=1\columnwidth]{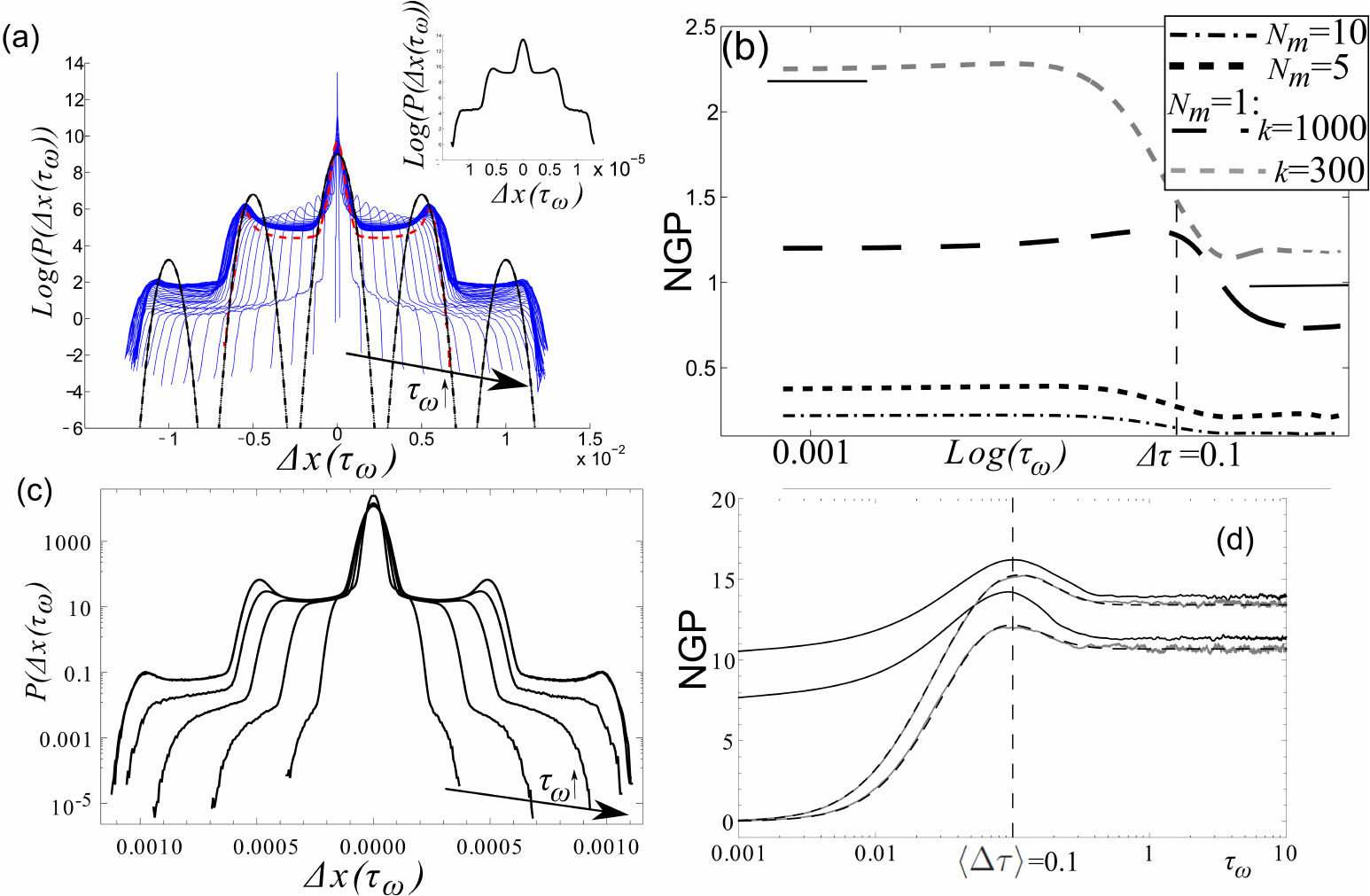}
\caption{(a) Distribution of particle displacements $P(\Delta x(\tau_{\omega}))$, for various lag time duration $\tau_{\omega}$ (blue lines), for a single motor and constant burst duration. The traces correspond to increasing time lag durations (black arrow), in the range: $1>\tau_{\omega}>5\times10^{-4}$. The red dashed line denotes the spatial distribution $P(x)$, and the black dashed lines denotes $P(\Delta x(\infty))$ (Eq.\ref{pdxinfinity}) for the approximation of $P(x)$ as a sum of three Gaussians. Inset shows the displacement distribution for very short lag times $\tau_{\omega}$. Parameters as in Fig.\ref{Fig2}c. (b) Calculated NGP for the $P(\Delta
x(\tau_{\omega}))$, for various number of motors ($N_m$), and confinement strength. The short horizontal black lines denote: (left) the NGP of $P(x)$, and (right) of $P(\Delta x(\infty))$ (Eq.\ref{pdxinfinity}), for the $k=1000,N_m=1$ case. (c) Displacement distributions (as in (a)) for a calculation without the inertia term (Eq.\ref{vdotdamp}, using: $k=1000$, Poissonian $\langle\Delta\tau\rangle=0.1$, and increasing lag time indicated by the arrow $\tau_{\omega}=10^{-3},2.5\times10^{-3},5\times10^{-3},10^-2,1$), and (d) the corresponding NGP, comparing the simulation (solid grey lines) to the analytical result (see Appendix for details, dashed lines), for $k=300,1000$ (top, bottom). The NGP with inertia is given by the solid black lines.}\label{Fig3}
\end{figure}

The distribution of relative particle displacements (Van Hove correlation function) $P(\Delta
x(\tau_{\omega}))$, where $\Delta
x(\tau_{\omega})=x(t+\tau_{\omega})-x(t)$ ($\tau_{\omega}$ is the lag-time duration), is a useful measure for the particle dynamics. We plot it in Fig.\ref{Fig3}a for the interesting regime of strong confinement and damping, and compared to the distribution of particle positions $P(x)$ (Fig.\ref{Fig2}c). We see that $P(\Delta
x(\tau_{\omega}))$ has double the number of peaks of $P(x)$, and is distinctly non-Gaussian for all $\tau_{\omega}$. In Fig.\ref{Fig3}c we show that the same qualitative behavior is obtained for Poissonian burst duration.

The deviations from Gaussianity are quantified in Fig.\ref{Fig3}b using the Non-Gaussianity Parameter (NGP) of the displacement distributions: $\kappa=\langle \Delta x^4\rangle/3\langle \Delta x^2\rangle^2-1$. This deviation of the kurtosis from the value for a Gaussian is an established measure for studying distributions \cite{rahman}. We find that the
NGP has a finite value for $\tau_{\omega}\rightarrow0$. This is as a consequence of the periods during which the
particle is accelerated by the active force, and the result is a
finite probability for displacements of the order of $\Delta x\simeq
v_0 \tau_{\omega}$ (inset of Fig.\ref{Fig3}a). With increasing $\tau_{\omega}$
the NGP reaches a maximum, at lag times that are of order $\Delta\tau$, where the full
effect of the active bursts is observed.

We find that the maximal value of the NGP for $P(\Delta
x(\tau_{\omega}))$ is close to the NGP of $P(x)$ (Fig.\ref{Fig3}b), which is a function for which we have a good analytic
approximation \cite{eyalPRL} (Eqs.\ref{ngppx},\ref{ngpmax}). In the limit of $\tau_{\omega}\rightarrow\infty$ the calculated NGP remains finite, and can be calculated analytically since the displacement distribution becomes
\begin{equation}
P(\Delta
x(\infty))=\int_{-\infty}^{\infty}P(x)P(x+\Delta
x) dx
\label{pdxinfinity}
\end{equation}
and the $P(x)$ in this regime is well approximated by the sum of shifted thermal Gaussians (inset of Fig.\ref{Fig2}c). This calculation fits well the simulated result (Fig.\ref{Fig3}b). For a larger number of motors, the distribution $P(\Delta x(\tau_{\omega}))$ approach a Gaussian (Fig.\ref{Fig3}b,S4).

In the limit where we discard inertia from the equations of motion (Eq.\ref{vdotdamp}), we can calculate the NGP analytically (see details in Appendix). In Fig.\ref{Fig3}c we plot the displacement distributions for this case, and in Fig.\ref{Fig3}d we show that indeed the analytical calculation describes exactly the simulation results. We find that this treatment captures correctly the qualitative features of the full system, such as the position of the peak, followed by a constant value at long lag times. The large discrepancy is in the limit of $\tau_{\omega}\rightarrow0$, where the inertial effects of the oscillations inside the trap are missing from Eq.\ref{vdotdamp}.

\section{Results: FDT}

An alternative method to characterize the non-equilibrium dynamics is through the deviations from the FDT \cite{Mackintosh}. We can quantify these deviations by defining an effective temperature, using the Fourier-transform of the position fluctuations ($S_{xx}(\omega)$) and linear response (susceptibility of the position to an external force $\chi(\omega)$) of the system. We can calculate both for our trapped particle position using Eq.(\ref{xeq}) for the $\tilde{\lambda}\Delta\tau\gg1$ limit, to get (for Poissonian burst duration $\Delta\tau$, see details in Appendix, Eqs.\ref{tfdtres}-\ref{tfdt})
\begin{eqnarray}
T_{FDT}(\omega)&=&\frac{\omega S_{xx}(\omega)}{2{\rm
Im}(\chi(\omega))}=\frac{N_m p_{on} f_0^2
\langle\Delta\tau\rangle}{\lambda (1+(\omega
\langle\Delta\tau\rangle)^2) } + T
\label{tfdt}
\end{eqnarray}
Note that $T_{FDT}(\omega)$ is independent of the shape of the
harmonic potential ($k$), and is identical to the result for a free
active particle \cite{eyalPRL}. This result highlights the fact that while different "effective temperatures" in an active system ($T_v$ and $T_x$, Eqs.\ref{txt},\ref{tx}) give a measure of the activity, they can have very different properties.

\section{Discussion}

We now use our results to interpret several experiments on
active gels in-vitro, and extract the values that characterize these active systems. In \cite{Mackintosh}
the break-down of the FDT was measured. Comparing to our $T_{FDT}$
(Eq.\ref{tfdt}) we find that the onset of the deviation from equilibrium
occurs for frequencies $\omega\leq\Delta\tau^{-1}$,
from which we find that: $\Delta\tau\approx100$ msec,
which is the scale of the release time of the myosin-II-induced stress \cite{Mackintosh} in this system. The measured deviation from the FDT was found to increase with decreasing frequency \cite{Mackintosh}, and
at the lowest measured frequencies the ratio was found to be
$T_{FDT}(\omega\rightarrow0)/T\approx20-100$. This number fixes for us
the combination of the parameters given in Eq.(\ref{tfdt}).

Recent experiments shed more detail on the active motion in this
system \cite{Schmidt}, and it was found that the tracer particle
performs random confined motion interspersed by periods of large
excursions. The confined motion part can be directly related to the mean-square displacement in our model $T_x$ (Eq.\ref{tx}), and is observed to
be a factor of $T_x/T\approx10-50$ larger than in the inert system (not
containing myosins)\cite{Schmidt}. These values are in general
agreement with the values extracted above for $T_{FDT}$ from \cite{Mackintosh}, and note
that we predict (Eqs.\ref{tx},\ref{tfdt}):
$T_{FDT}(\omega\rightarrow0)/T_x=1+\tilde{\lambda}\langle\Delta\tau\rangle>1$.

Furthermore, in these experiments \cite{Schmidt} it was observed
that the distribution of relative particle displacements $P(\Delta
x(\tau_{\omega}))$ is highly non-Gaussian.
Comparing to Fig.\ref{Fig3}b we note that
both the experiments and in our calculations the
NGP has a finite value for $\tau_{\omega}\rightarrow0$. With increasing $\tau_{\omega}$
the NGP reaches a maximum, both in the experiments and in our
calculations (Fig.\ref{Fig3}b,d). By comparing to our model we expect the peak to appear at $\tau_{\omega}\approx\Delta\tau$, so the observations \cite{Schmidt} suggest the burst duration is of order $\Delta\tau\approx1-10$sec, in agreement with similar studies \cite{Koenderink2012pre,konderink2014}. Note that very similar NGP time-scales were observed in living cells \cite{fredberg2005,Gal2013} Our model predicts that the maximal value of the NGP is a non-monotonous function of $p_{on}$, and this may be explored by varying the concentration of ATP in the system. Furthermore, from our model we predict that the NGP decrease with decreasing active force, and increasing stiffness of the confining network (Fig.\ref{Fig3}b,d, Eq.\ref{ngpmax}). These predictions can be related to the observed activity-dependence of the NGP in cells \cite{Gal2013}, and the decay of the NGP during the aging and coarsening of an active gel \cite{Koenderink2012pre}.

The large observed deviations from Gaussianity indicate that the particle is in the
strong confinement regime: $\tilde{\lambda}\langle\Delta\tau\rangle>1$. The maximal value of the observed NGP $\approx2-4$ can be used to get an estimate of $T_x$, by taking it to be equal to $NGP_{max}$ (Eq.\ref{ngpmax}). This gives us: $T_x\approx10-30 k_{B}T$ and $p_{on}\approx2-3\%$. This value of $T_x$ is in good agreement with the estimate made above. The value of $p_{on}$ is in agreement with the observation that the waiting-time between bursts is much longer than the burst duration \cite{Mackintosh}, and with the measured duty-ratio of myosin-II \cite{howard}.

In the limit of $\tau_{\omega}\rightarrow\infty$ the observed NGP of the displacement distribution $P(\Delta x(\tau_{\omega}))$
decays to zero \cite{Schmidt}, while for the calculated confined particle the NGP remains finite (Fig.\ref{Fig3}b,d). At long times ($\gtrapprox10$sec) the observed trajectory has large excursions \cite{Schmidt}, which we
interpret as the escape of the particle from the confining potential. The ensuing
hopping-type diffusion, causes the NGP to vanish, as for free diffusion \cite{manning}. Within our model we therefore interpret the observed time-scale of the vanishing of the NGP, $\tau_{\omega}\approx10-100$ sec, as the time-scale which corresponds to the mean trapping time of the bead within the confining actin gel. Beyond this time-scale the bead has a large chance to escape the confinement, and hop to a new trapping site, which corresponds to a re-organization of the actin network. The real actin-myosin gel undergoes irreversible processes that make its properties time-dependent and render it inhomogeneous \cite{Koenderink2012pre,Bernheim,konderink2011,Bausch}. Such effects make the comparison to the model much more challenging. Large deviations from Gaussianity were also observed for the Van Hove correlations in other forms of active gels \cite{activeDNA}.

\section{Conclusion}

We investigated here the dynamics of a trapped active particle, with several interesting results: (i) The activity leads to strong deviations from equilibrium, such as the break-down of the virial theorem and equipartition. We find that in the presence of elastic restoring forces the activity is mostly "stored" in the potential energy of the system. (ii) Different "effective temperatures" give a measure of the activity, and some are dependent on the stiffness of the elastic confinement. (iii) The displacement, position and velocity distributions of the particle are highly non-Gaussian in the regime of strong elastic confinement and small number of dominant motors. These distributions can be used, together with our simple model, to extract information about the microscopic properties of the active motors. Note that in our model the activity affects the motion and position distributions of the trapped particle, which is complimentary to models where the activity drives only the large-scale reorganization that moves the particle between trapping sites \cite{epl2015,kanazawa}, or leads to network collapse \cite{sheinman2015}. The results of this model are in good agreement with observations of the dynamics of tracer beads inside active gels, and the simplicity of the model may make it applicable for a wide range of systems. More complex visco-elastic relations can be used in place of the simple elasticity presented here, to describe the dynamics inside living cells \cite{Wilhelm,Gallet2009}, as well as non-linear elasticity \cite{nonlinear}. Note that in most current experiments on actin-myosin gels, the myosin-driven activity is strong enough to lead to large-scale reorganization of the actin network, eventually leading to the network collapse \cite{Koenderink2012pre,Bernheim,konderink2011,Bausch}. In order to observe the active motion for the elastically-trapped tracer in the intact network, which we have calculated, much weaker active forces will be needed. Our work can therefore give motivation for such future studied.

\widetext
\appendix


\section{Numerical simulations}

The simulations of the dynamics of the particle inside the 1D harmonic potential were carried out using explicit Euler integration of Eq.\ref{vdot}. We were careful to use a small time-step $\Delta t$, such that it was always an order of magnitude smaller than the smallest time-scale in the problem. The time-scales in the problem are: $\tau$,$\Delta\tau$ and $\sqrt{2/k}$, where the last time-scale is that of the oscillation frequency of the particle inside the harmonic potential.

The iterative equations take the following form in terms of the sampling time $\Delta t$
\begin{subequations}
\begin{eqnarray}
v(t+\Delta t) &=& v(t) + \left[ - \lambda v(t) - k x(t) + f_a(t) \right] \Delta t  +  \sqrt{2\lambda T \Delta t} \eta,\label{A1}\\
x(t+\Delta t) &=& x(t) + v(t)\Delta t, \label{A2}
\end{eqnarray}
\end{subequations}
where $\eta$ is a random Gaussian variable with zero mean and variance $1$.
Considering that both the waiting time and the persistence time are exponentially distributed with mean values $\tau$ and $\Delta \tau$, respectively, the iterative equation for the active force $f_a$ obeys
\begin{equation}
f_a(t+\Delta t) =
\begin{cases}
f_a(t) & \text{if} \quad f_a(t)\neq 0 \quad \text{prob.} \quad 1-\Delta t/\Delta\tau
\,\,,
\\
f_a(t)  & \text{if} \quad f_a(t)=0 \quad \text{prob.} \quad 1-\Delta t/\tau
\,\,,
\\
0 & \text{if} \quad f_a(t)\neq 0 \quad \text{prob.} \quad \Delta t/\Delta\tau
\,\,,
\\
\epsilon_{\{-f,f\}} & \text{if} \quad f_a(t)=0 \quad \text{prob.} \quad \Delta t/\tau
\,\,,
\end{cases}
\end{equation}
where $\epsilon_{\{-f,f\}}=\{f,-f\}$ with same probability.

\section{Position fluctuations of a trapped particle}

From the model equations of motion (Eq.\ref{vdot}), we can calculate the mean-square fluctuations in the particle position for a shot-noise force correlations with average burst duration $\Delta\tau$.
We begin by Fourier transforming Eq.\ref{vdot} to get
\begin{eqnarray}
-\omega^2\widetilde{x}&=&i\omega\lambda \widetilde{x} +\widetilde{f_a}+\widetilde{f_T}-k \widetilde{x}
\label{vdotFT}
\end{eqnarray}
where the $\widetilde{}$ denotes the FT. From Eq.\ref{vdotFT} we get
\begin{eqnarray}
\widetilde{x}(\omega)=\frac{\widetilde{f_a}(\omega)+\widetilde{f_T}(\omega)}{-\omega^2-i\omega\lambda+k}
\label{xFT}
\end{eqnarray}
The fluctuations (correlations) are therefore
\begin{eqnarray}
\langle x^2\rangle(\omega)&=&\langle\widetilde{x}(\omega)\widetilde{x}^*(\omega)\rangle=\frac{\langle\widetilde{f_a^2}\rangle(\omega)+\langle\widetilde{f_T^2}\rangle(\omega)}{\left(k-\omega^2\right)^2+(\omega\lambda)^2} \label{fullxFT}
\end{eqnarray}
where we have: $\langle\widetilde{f_a^2}\rangle(\omega)=N_mp_{on}f_0^2\frac{\Delta\tau}{1+(\omega\Delta\tau)^2}$ (Poissonian shot-noise with mean burst length $\Delta\tau$), and $\langle\widetilde{f_T^2}\rangle(\omega)=2\lambda T$ (thermal white noise) \cite{eyalPRL}.

For the active part alone, we get
\begin{equation}
\langle x^2\rangle=\frac{N_mp_{on}f_0^2}{2\pi}\int_0^{\infty}\frac{1}{(\omega^2-k)^2+(\omega\lambda)^2}\frac{\Delta\tau}{1+(\omega\Delta\tau)^2}d\omega
\label{fullxsol}
\end{equation}
The solution for this integral is quite lengthy.
In the limit of weak trapping, $k\rightarrow0$, we get that the mean-square displacement diverges
\begin{equation}
\langle x^2\rangle\rightarrow 2\frac{N_mp_{on}f_0^2\Delta\tau}{k\left(k\Delta\tau+\lambda\right)}
\label{x2weaktraplimit}
\end{equation}
such that the mean potential energy in this limit approaches a constant value
\begin{equation}
T_x'\simeq k\langle x^2\rangle\rightarrow 2\frac{N_mp_{on}f_0^2\Delta\tau}{\left(k\Delta\tau+\lambda\right)}
\label{txweak}
\end{equation}

\begin{figure}
\includegraphics[width=6in]{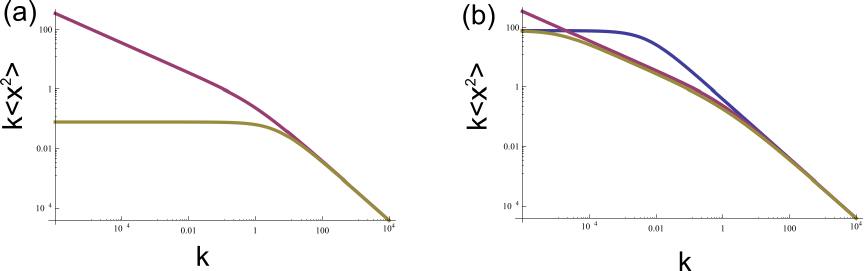}
\caption{Calculated mean-square position fluctuations (plotted as a mean potential energy) for the trapped particle: Brown line- full solution, purple line- approximate solution (Eq.\ref{largekxsol}), blue line- approximate expression $T_x'$ for the limit $\lambda \Delta \tau\gg1$ (Eq.\ref{txA}). In both panels we used $\Delta\tau=1$, and (a) $\lambda=10$, (b) $\lambda=0.01$. In (a) the blue line agrees perfectly with the full solution, while in (b) it has a discrepancy at intermediate confinements.}\label{figx1}
\end{figure}

In the limit of large $k$, we can expand the integrand of Eq.(\ref{fullxsol}) in powers of $k^{-1}$ to get this integral
\begin{equation}
\langle x^2\rangle=\frac{2N_mp_{on}f_0^2}{\pi}\int_0^{\sqrt{k}}\frac{\Delta\tau}{k^2(1+(\omega\Delta\tau)^2)}d\omega
\label{fullxsolk}
\end{equation}
which is also bound with a maximal frequency corresponding to the natural frequency of the harmonic trap. This integral gives a simple expression, which gives a good fit description as long as $k\gg\lambda^2$ (Fig.\ref{figx1})
\begin{equation}
\langle x^2\rangle_{k}=\frac{2N_mp_{on}f_0^2}{\pi k^2}\arctan{[\sqrt{k/2}\Delta\tau]}
\label{largekxsol}
\end{equation}
Finding the value of $k$ for which the scaling changes from $\langle x^2\rangle\sim k^{-3/2}$ to $\langle x^2\rangle\sim k^{-2}$, is simply by equating the large and small $k$ limits of $\langle x^2\rangle_{k}$ (Eq.\ref{largekxsol}).

In the limit of $\lambda \Delta\tau\gg1$ we find the simple approximate expression (Fig.\ref{figx1})
\begin{equation}
T_x'\simeq k\langle x^2\rangle\simeq\frac{\Delta\tau N_mp_{on}f_0^2}{8\lambda^2(k\Delta\tau/2\lambda+1)}
\label{txA}
\end{equation}

The numerical simulations, in the highly damped limit ($\lambda \Delta\tau\gg1$) indicate the $k^{-1}$ and $k^{-2}$ limits (Fig.\ref{simSquare}a).

\section{Velocity fluctuations of a trapped particle}

Similar to the procedure for the position fluctuations described above, we can calculate the velocity fluctuations. The mean-square fluctuations in the particle velocity are given simply from Eq.(\ref{fullxsol}) by
\begin{equation}
\langle v^2\rangle=\frac{N_mp_{on}f_0^2}{2\pi}\int_0^{\infty}\frac{\omega^2}{(\omega^2-k)^2+(\omega\lambda)^2}\frac{\Delta\tau}{1+(\omega\Delta\tau)^2}d\omega
\label{fullvsol}
\end{equation}
The solution for this integral is again quite lengthy.
As for the position distribution, we can find an approximation for the large $k$ limit, using
   \begin{equation}
\langle v^2\rangle=\frac{2N_mp_{on}f_0^2}{\pi}\int_0^{\sqrt{k}}\frac{\omega^2\Delta\tau}{k^2(1+(\omega\Delta\tau)^2)}d\omega
\label{fullvsolk}
\end{equation}
which is also bound with a maximal frequency corresponding to the natural frequency of the harmonic trap. This integral gives a simple expression, which gives a good fit description as long as $k\gg\lambda^2$
\begin{equation}
\langle v^2\rangle_{k}=\frac{2N_mp_{on}f_0^2}{\pi k^2\Delta\tau^2}\left(\sqrt{k/2}\Delta\tau-\arctan{[\sqrt{k/2}\Delta\tau]}\right)
\label{largekvsol}
\end{equation}

\begin{figure}
\includegraphics[width=6in]{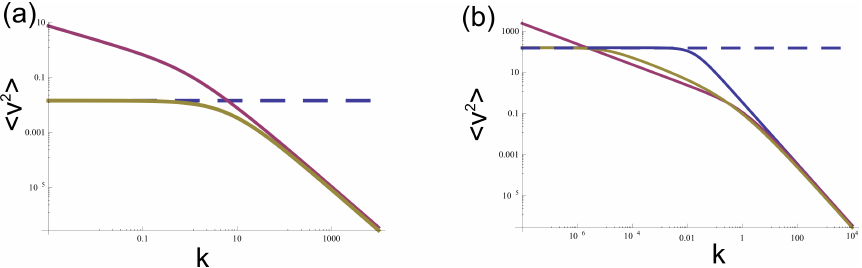}
\caption{Calculated mean-square velocity fluctuations for the trapped particle: Brown line- full solution, purple line- approximate solution (Eq.\ref{largekvsol}), blue line- highly damped limit (Eq.\ref{tvapprox}), and the dashed blue line is the free-particle value \cite{eyalPRL}. In both panels we used $\Delta\tau=1$, and (a) $\lambda=10$, (b) $\lambda=0.01$.}\label{figx2}
\end{figure}

The scaling of $\langle v^2\rangle_{k}$  changes from $\langle v^2\rangle\sim k^{-0.5}$ to $\langle v^2\rangle\sim k^{-3/2}$ as $k$ increases (Fig.\ref{figx2}b).

In the limit of $\lambda \Delta\tau\gg1$ we have the simple approximate expression
\begin{equation}
\langle v^2\rangle\simeq\frac{\Delta\tau N_mp_{on}f_0^2}{4(\lambda(1+\Delta\tau\lambda)+\pi\Delta\tau^2\sqrt{k^3/8})}
\label{tvapprox}
\end{equation}
which fits quite well the full expression in Fig.\ref{figx2}a.

\begin{figure}
\includegraphics[width=6in]{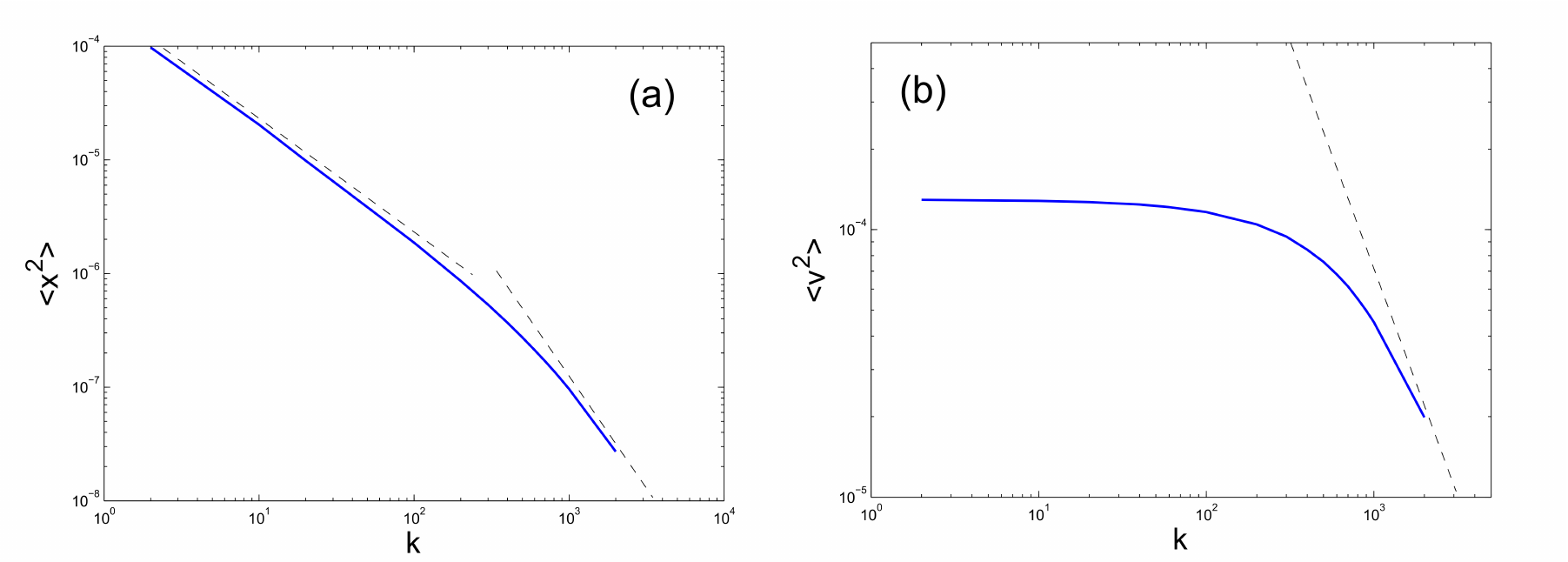}
\caption{Simulated mean-square particle displacements (a) and velocity (b) in the limit of $\lambda \Delta \tau\gg1$, using: $\Delta\tau=1$, $\lambda=50$, $f_0=1$, $p_{on}=0.1$. The dashed lines indicate the power-laws with exponents $-1,-2$ and $-3/2$ in (a) and (b) respectively.}\label{simSquare}
\end{figure}

The numerical simulations, in the highly damped limit ($\lambda \Delta\tau\gg1$) indicate the $k^{0}$ and $k^{-3/2}$ limits (Fig.\ref{simSquare}b).

\section{Effective temperature due to forces from distant (and numerous) motors}

In a linear elastic medium, the displacements and stresses decay from a point source (at least) as $1/r^2$. Since there are numerous distant motors affecting the bead, their cumulative random forces are most likely to give rise to Gaussian distribution of position and velocities for the trapped particle. Each shell (of thickness $dr$) at radius $r$ from the tracer beads has $N_m(r)=4\pi r^2 \rho dr$ motors (at constant density $\rho$), and therefore they contribute to the mean-square velocity the following contribution (in the limit of $\lambda \Delta\tau\gg1$, using Eq.\ref{tvapprox})
\begin{equation}
\langle v^2\rangle\simeq N_m(r)\left(N_mp_{on}f_0\frac{a^2}{r^2}\right)^2\frac{\Delta\tau}{4(\lambda(1+\Delta\tau\lambda)+\pi\Delta\tau^2\sqrt{k^3})}\propto\frac{1}{r^2}
\label{tvapproxfar}
\end{equation}
where we isolated the number of motors and the $r$-dependence of the active forces, and introduced a length-scale $a$ beyond which the far-field calculation holds. Integrating this expression we get
\begin{equation}
\langle v^2\rangle_{far}\simeq \langle v^2\rangle_0\left(4\pi\rho a^3\right)
\label{tvapproxfar1}
\end{equation}
where $\langle v^2\rangle_0$ is the value for the single proximal motor given in Eq.\ref{tvapprox}. We find that the far-field contribution of the distant motors is proportional to their density $\rho$.

%
%

\section{Displacement distribution for numerous motors}

\begin{figure}
\includegraphics[width=6in]{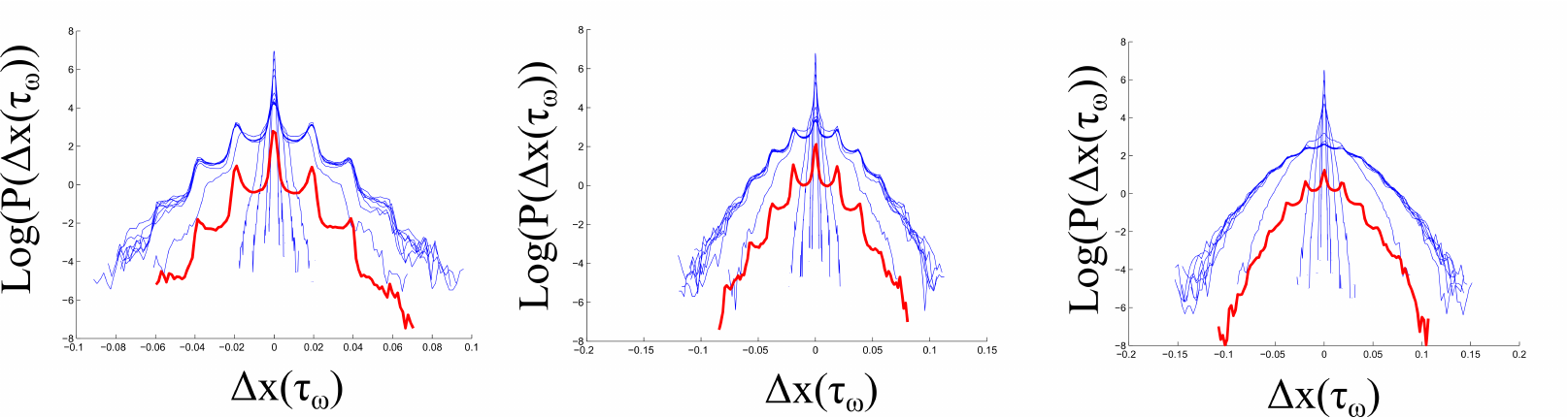}
\caption{Simulated particle position distribution $P(x)$ (red lines) and displacement distributions $P(\Delta
x(\tau_{\omega}))$ (blue lines), for increasing number of motors: $N_m=5,10,20$ (left to right), using: $\Delta\tau=1$, $\lambda=50$, $f_0=1$, $p_{on}=0.1$.}\label{DisplacementNm}
\end{figure}

As the number of motors kicking the particle ($N_m$) increases, we find that the distribution of the particle position becomes more Gaussian, even in the limit of larger damping $\lambda\Delta\tau\gg1$ and strong confinement $\tilde{\lambda}\Delta\tau\gg1$. We demonstrate this in Fig.\ref{DisplacementNm}, which shows that the position distributions $P(x)$ and the displacement distributions $P(\Delta x(\tau_{\omega}))$ approach a Gauassian for $N_m$ larger than $\sim10$.

\section{NGP for the highly damped limit}

We find that the maximal value of the NGP for $P(\Delta
x(\tau_{\omega}))$ is close to the NGP of $P(x)$ (Fig.\ref{Fig3}b), which is a function for which we have a good analytic
approximation \cite{eyalPRL}, given by (for a single motor)
\begin{equation}
NGP
(N_m=1)=\frac{4(1-3p_{on})p_{on}^3T_{x,1}^2}{3(1+2p_{on}^2T_{x,1})^2}
\label{ngppx}
\end{equation}
where $T_{x,1}$ is the effective temperature of the spatial
distribution (Eq.\ref{txt},\ref{tx}) for $p_{on}=1$. The maximal value of the NGP for a single motor, as a function of $p_{on}$ is obtained from Eq.\ref{ngppx}, at: $p_{on}=\alpha/(2+6\alpha)$, and is given by
\begin{equation}
NGP_{max}=\frac{(1+3\alpha)^2}{3\alpha(2+3\alpha)}-1
\label{ngpmax}
\end{equation}
where $\alpha=k_{B}T/T_{x,1}$. This is a monotonously decreasing function of the stiffness $k$, due to the decrease in $T_{x,1}$ in stiffer gels (Eqs.4,5).

\section{Effective temperature from the FDT, $T_{FDT}$}

Following \cite{eyalPRL}, and using Eq.\ref{xeq}, we can write for the $\tilde{\lambda}\Delta\tau\gg1$ limit (when $T=0$)
\begin{eqnarray}
{\rm Response:}&& \chi_{xx}(\omega)=\frac{1}{\gamma(i\omega -
\tilde{\lambda})} \label{tfdtres}\\
{\rm Fluctuations:}&&
S_{xx}(\omega)=\frac{f_0^2}{\lambda(\tilde{\lambda}^2+\omega^2)}\frac{\langle\Delta\tau\rangle}{1+(\omega
\langle\Delta\tau\rangle)^2}\label{tfdtfluct}\\
\Rightarrow T_{FDT}(\omega)&=&\frac{\omega S_{xx}(\omega)}{2{\rm
Im}(\chi(\omega))}=\frac{N_m p_{on} f_0^2
\langle\Delta\tau\rangle}{\lambda (1+(\omega
\langle\Delta\tau\rangle)^2)}
\label{tfdt}
\end{eqnarray}
resulting in Eq.\ref{tfdt}.

\section{Analytic calculation of the NGP without inertia}

To compute the expression of the NGP, we derive the mean quartic displacement (MQD) $\avg{\Delta x^4}$ in the regime where it is time translational invariant
\begin{equation}\label{eq:ngp}
\avg{\Delta x^4}=\avg{\Delta x_\text{\tiny T}^4} + \avg{\Delta x_\text{\tiny A}^4} + 6\avg{\Delta x_\text{\tiny T}^2} \avg{\Delta x_\text{\tiny A}^2}
\,\,,
\end{equation}
where the subscripts T and A refer respectively to the thermal and active contributions. The expression of the MSD is given by
\begin{equation}
\avg{\Delta x_\text{\tiny T}^2}(t) = \frac{2\kb T}{k}\pnt{1-\ee^{-t/\td}}\,,\quad \avg{\Delta x_\text{\tiny A}^2}(t) = \frac{2\kb \Ta/k}{(\tau/\td)^2-1}\brt{ \frac{\tau}{\td} \pnt{ 1-\ee^{-t/\tau} } +\ee^{-t/\td}-1 }
\,\,,
\end{equation}
where  $\td=\lambda/k$ is a thermal relaxation time scale.
The MQD under purely thermal conditions is related to the thermal MSD since the thermal process is Gaussian $\avg{\Delta x_\text{\tiny T}^4}=3\avg{\Delta x_\text{\tiny T}^2}^2$. To compute the active MQD, we separate the position displacement $\Delta x_\text{\tiny A}(\ti,\tf)=x_\text{\tiny A}(\tf)-x_\text{\tiny A}(\ti)$ in several contributions, such that $\avg{\Delta x_\text{\tiny A}^4}$ is a power law combination of these contributions. We compute each term using the active force statistics, and take the limit of large $\ti$ at fixed $t$ corresponding to the time translational regime. The advantage of the separation we propose is that each term of the active MQD converges in such limit. The appropriate separation is
\begin{subequations}
\begin{eqnarray}
\Delta x_\text{{\tiny A},a}(\ti,\tf) &=& \pnt{\ee^{-t/\td}-1}\int\limits^{\ti} \dd t' \chi(\ti-t')\fa(t')
\,\,,
\\
\Delta x_\text{{\tiny A},b}(\ti,\tf) &=& \int\limits^t \dd t' \chi(t-t')\fa(\ti+t')
\,\,,
\end{eqnarray}
\end{subequations}
where $\chi(t)=\ee^{-t/\td}/\lambda$ is the non--causal response function, and $t=\tf-\ti$ is the time lag. In the time translational regime, we compute
\begin{subequations}
\begin{eqnarray}
\avg{\Delta x_\text{{\tiny A},a}^4}(t) &=& \Ta^2\frac{3 \tau_\text{r} ^4 (2 \tau_0 +\tau_\text{r} ) (\tau_0 +\tau ) e^{-\frac{4 t}{\tau_\text{r} }} \left(e^{t/\tau_\text{r} }-1\right)^4}{\lambda ^2 (\tau_\text{r} +\tau ) (\tau_\text{r} +3 \tau ) (\tau_\text{r}  (\tau_0 +\tau )+2 \tau_0  \tau )}
\,\,,
\\
\avg{\Delta x_\text{{\tiny A},a}^3 \Delta x_\text{{\tiny A},b}} (t) &=& \Ta^2\frac{3 \tau_\text{r} ^4 \tau  (2 \tau_0 +\tau_\text{r} ) (\tau_0 +\tau ) e^{-\frac{4 t}{\tau_\text{r} }} \left(e^{t/\tau_\text{r} }-1\right)^3 \left(e^{t \left(\frac{1}{\tau_\text{r} }-\frac{1}{\tau }\right)}-1\right)}{\lambda ^2 (\tau_\text{r} -\tau ) (\tau_\text{r} +\tau ) (\tau_\text{r} +3 \tau ) (\tau_\text{r}  (\tau_0 +\tau )+2 \tau_0  \tau )}
\,\,,
\\
\avg{\Delta x_\text{{\tiny A},a}^2 \Delta x_\text{{\tiny A},b}^2}(t) &=& \Ta^2\frac{\tau_\text{r} ^4 \left(e^{t/\tau_\text{r} }-1\right)^2 e^{-\frac{4 t}{\tau_\text{r} }-\frac{t}{\tau }} }{\lambda ^2 (\tau_\text{r} -\tau_0 ) (\tau -\tau_\text{r} ) (\tau_\text{r} +\tau )^2 (\tau_\text{r}  (\tau_0 +\tau )-2 \tau_0  \tau ) (\tau_\text{r}  (\tau_0 +\tau )+2 \tau_0  \tau )}
\nonumber
\\
& &\times\bigg(4 \tau_0 ^4 (\tau -\tau_\text{r} ) (\tau_\text{r} +\tau ) e^{\frac{2 t}{\tau_\text{r} }-\frac{t}{\tau_0 }}+(\tau_\text{r} -\tau_0 ) (\tau -\tau_\text{r} ) \left(\tau_\text{r} ^2 (\tau_0 +\tau )^2-4 \tau_0 ^2 \tau ^2\right) e^{t \left(\frac{2}{\tau_\text{r} }+\frac{1}{\tau }\right)}
\nonumber
\\
& &+(\tau_0 -\tau_\text{r} ) (\tau_0 +\tau ) (\tau_\text{r} +\tau ) e^{t/\tau } \left(4 \tau_0 ^2 \tau -\tau_\text{r} ^2 (\tau_0 +\tau )\right)
\nonumber
\\
& &-2 (\tau_0 +\tau ) e^{t/\tau_\text{r} } \left(2 \tau_0 ^2 \tau_\text{r} +\tau  (\tau_0 -\tau_\text{r} ) (2 \tau_0 +\tau_\text{r} )\right) (2 \tau_0  \tau -\tau_\text{r}  (\tau_0 +\tau ))\bigg)
\,\,,
\\
\avg{\Delta x_\text{{\tiny A},a}\Delta x_\text{{\tiny A},b}^3} (t) &=& -\Ta^2\frac{3 \tau_\text{r} ^4 e^{-\frac{4 t}{\tau_\text{r} }} \left(e^{t/\tau_\text{r} }-1\right)}{\lambda ^2 \tau  (\tau_\text{r} +\tau )}
\bigg(\frac{\tau ^2 (2 \tau_0 -\tau_\text{r} ) (\tau_0 +\tau )}{(\tau_\text{r} -3 \tau ) (\tau_\text{r} -\tau ) (\tau_0  (\tau_\text{r} -2 \tau )+\tau_\text{r}  \tau )}
\nonumber
\\
& &+\frac{2 \tau_0 ^4 e^{-t \left(\frac{1}{\tau_0 }-\frac{2}{\tau_\text{r} }+\frac{1}{\tau }\right)}}{(\tau_0 -\tau_\text{r} ) (\tau_0 +\tau_\text{r} ) (\tau_\text{r}  (\tau_0 +\tau )-2 \tau_0  \tau )}+\frac{(\tau_0 +\tau ) (\tau_0  (\tau_\text{r} +\tau )-\tau_\text{r}  \tau ) e^{t \left(\frac{1}{\tau_\text{r} }-\frac{1}{\tau }\right)}}{(\tau_\text{r} -\tau_0 ) \left(\tau_\text{r} ^2-\tau ^2\right)}
\nonumber
\\
& &+\frac{(\tau_0 +\tau ) (\tau  (\tau_0 +\tau_\text{r} )-\tau_0  \tau_\text{r} ) e^{\frac{3 t}{\tau_\text{r} }-\frac{t}{\tau }}}{(\tau_0 +\tau_\text{r} ) \left(\tau_\text{r} ^2-4 \tau_\text{r}  \tau +3 \tau ^2\right)}+\frac{\tau ^2 e^{\frac{2 t}{\tau_\text{r} }}}{\tau_\text{r} ^2-\tau ^2}\bigg)
\,\,,
\\
\avg{\Delta x_\text{{\tiny A},b}^4} (t) &=& \Ta^2\frac{3 \tau_\text{r} ^4 e^{-\frac{4 t}{\tau_\text{r} }} }{\lambda ^2 \tau  (\tau_\text{r} +\tau )}
\bigg(-\frac{\tau  (2 \tau_0 -\tau_\text{r} ) (\tau_0 +\tau ) (\tau_\text{r} +\tau )}{(\tau_\text{r} -3 \tau ) (\tau_\text{r} -\tau ) (\tau_0  (\tau_\text{r} -2 \tau )+\tau_\text{r}  \tau )}
\nonumber
\\
& &+\frac{8 \tau_0 ^5 (\tau_\text{r} +\tau ) e^{t \left(-\frac{1}{\tau_0 }+\frac{2}{\tau_\text{r} }-\frac{1}{\tau }\right)}}{(\tau_\text{r} -\tau_0 ) (\tau_0 +\tau_\text{r} ) (\tau_\text{r}  (\tau_0 +\tau )-2 \tau_0  \tau ) (\tau_\text{r}  (\tau_0 +\tau )+2 \tau_0  \tau )}
\nonumber
\\
& &-\frac{4 (\tau_0 +\tau ) (\tau  (\tau_0 +\tau_\text{r} )-\tau_0  \tau_\text{r} ) e^{\frac{3 t}{\tau_\text{r} }-\frac{t}{\tau }}}{(\tau_0 +\tau_\text{r} ) \left(\tau_\text{r} ^2-4 \tau_\text{r}  \tau +3 \tau ^2\right)}+\frac{4 (\tau_0 +\tau ) (\tau_0  (\tau_\text{r} +\tau )-\tau_\text{r}  \tau ) e^{t \left(\frac{1}{\tau_\text{r} }-\frac{1}{\tau }\right)}}{(\tau_0 -\tau_\text{r} ) (\tau_\text{r} -\tau ) (\tau_\text{r} +3 \tau )}
\nonumber
\\
& &+\frac{\tau  (2 \tau_0 +\tau_\text{r} ) (\tau_0 +\tau ) e^{\frac{4 t}{\tau_\text{r} }}}{(\tau_\text{r} +3 \tau ) (\tau_\text{r}  (\tau_0 +\tau )+2 \tau_0  \tau )}+\frac{2 \tau  e^{\frac{2 t}{\tau_\text{r} }}}{\tau -\tau_\text{r} }\bigg)
\,\,,
\end{eqnarray}
\end{subequations}
from which we deduce
\begin{equation}
\avg{\Delta x_\text{\tiny A}^4} = \avg{\Delta x_\text{{\tiny A},a}^4} + 3 \avg{\Delta x_\text{{\tiny A},a}^4 \Delta x_\text{{\tiny A},b}} + 6 \avg{\Delta x_\text{{\tiny A},a}^2 \Delta x_\text{{\tiny A},b}^2} + 3 \avg{\Delta x_\text{{\tiny A},a}\Delta x_\text{{\tiny A},b}^3} + \avg{\Delta x_\text{{\tiny A},b}^4}
\,\,.
\end{equation}

\begin{acknowledgments}
N.S.G would like to thank the ISF grant 580/12 for support. This work is made possible through the historic generosity of the Perlman family.
\end{acknowledgments}

\end{document}